\documentclass{aa}

\usepackage{graphicx}
\usepackage{txfonts}

\begin{document}

\title{On the mass and metallicity distribution of parent AGB stars of presolar SiC}

\author{S. Cristallo
\inst{1,2}
\and A. Nanni
\inst{3,1}
\and G. Cescutti
\inst{4,5}
\and I. Minchev
\inst{6}
\and N. Liu
\inst{7,8}
\and D. Vescovi
\inst{9,2,1}
\and D. Gobrecht
\inst{10}
\and L. Piersanti
\inst{1,2}}

\institute{INAF, Osservatorio Astronomico d'Abruzzo, Via Mentore Maggini snc, 64100 Teramo, Italy
\and 
INFN, Sezione di Perugia, Via A. Pascoli snc, 06123 Perugia, Italy
\and
Aix Marseille Univ, CNRS, CNES, LAM, Marseille, France
\and INAF, Osservatorio Astronomico di Trieste, Via G.B. Tiepolo 11, I-34143 Trieste, Italy
\and IFPU, Istitute for the Fundamental Physics of the Universe, Via Beirut,  2, 34151, Grignano, Trieste, Italy
\and Leibniz Institut f\"ur Astrophysik Potsdam (AIP), An der Sterwarte 16, D-14482 Potsdam, Germany
\and
Laboratory for Space Sciences and Physics Department, Washington University in St. Louis, St. Louis, MO 63130, USA
\and McDonnell Center for the Space Sciences, Washington University in St. Louis, St. Louis, MO 63130, USA
\and Gran Sasso Science Institute, Viale Francesco Crispi, 7, 67100 L'Aquila, Italy
\and Institute of Astronomy, KU Leuven, Celestijnenlaan 200D, 3001 Leuven, Belgium}

\date{Received ; accepted } 

\abstract
{}{The vast majority ($\gtrsim$90\%) of presolar SiC grains identified in primitive meteorites are relics of ancient asymptotic giant branch (AGB) stars, whose ejecta were incorporated into the Solar System during its formation. Detailed characterization of these ancient stardust grains has revealed precious information on mixing processes in AGB interiors in great detail. However, the mass and metallicity distribution of their parent stars still remains ambiguous, although such information is crucial to investigating the slow neutron capture process, whose efficiency is mass- and metallicity-dependent.}{Using a well-known Milky Way chemo-dynamical model, we follow the evolution of the AGB stars that polluted the Solar System at 4.57 Gyr ago and weighted the stars based on their SiC dust productions.}{We find that presolar SiC in the Solar System predominantly originated from AGB stars with M $\sim$ 2 M$_\odot$ and Z $\sim$ Z$_\odot$. Our finding well explains the grain-size distribution of presolar SiC identified in situ in primitive meteorites. Moreover, it provides complementary results to very recent papers dealing with the characterization of parent stars of presolar SiC. 
%and represents a valid alternative to the recent suggestion of more massive and more metal-rich AGB stars as the parent stars of presolar SiC grains.
}{}


\maketitle

\section{Introduction}
\label{sec:intro}
\begin{figure}
\centering
\includegraphics[width=\columnwidth]{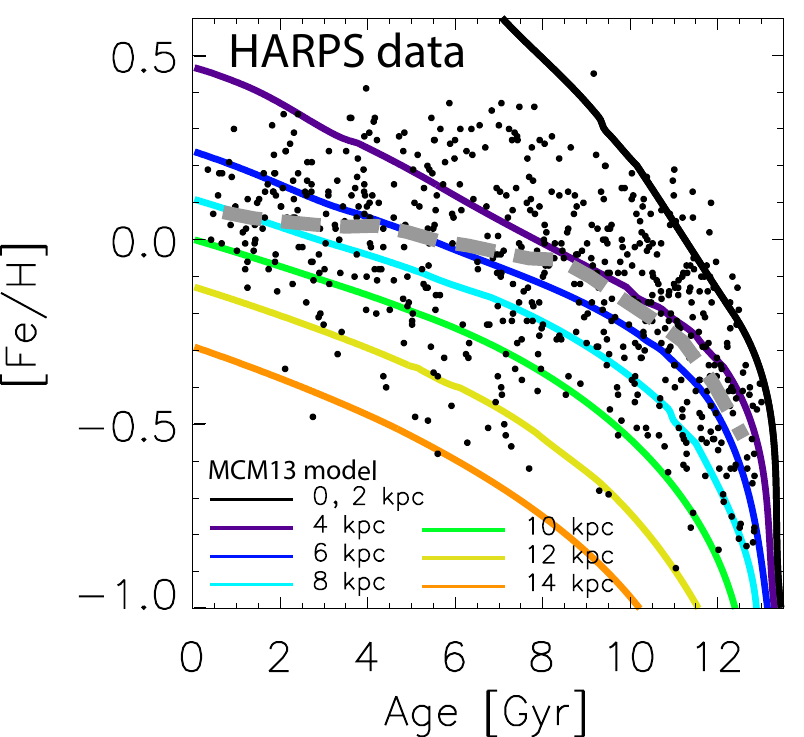}
\caption{Scatter points are HARPS CTO data (recently re-analyzed data by \citealt{delga17} and ages estimates by \citealt{ande18}), which represent the age metallicity relation (AMR) in the Galaxy. Dashed grey line shows the average of the total sample, which appears flat in the last $\sim$5 Gyr. Color-coded curves indicate averages of sub-samples according to their birth locations from the MCM13 model. The apparent flatness in the local AMR can be explained by the superposition of well-defined AMRs of stars born in narrow birth-radius bins.}
\label{fig:gce}
\end{figure} 
Stars are very efficient nuclear cauldrons, making the vast majority of chemical elements in the Universe. Isotopes freshly synthesized in their interiors are directly ejected  (in case of explosive events) into the interstellar medium (ISM) or carried to the surface by convective mixing episodes (known as Dredge Ups) and then lost to the ISM by stellar winds. In stars, elements heavier than iron are mostly produced via two neutron capture processes: the \textit{rapid} process (\textit{r}-process; see \citealt{cowan2020} for a review) and the \textit{slow} process (\textit{s}-process; see \citealt{bu99} for a review). The latter mainly occurs during the asymptotic giant branch (AGB) phase of low-mass stars (see, e.g., \citealt{stra06}). \\
\indent Isotopic ratios of {\it s}-process elements can be obtained in single $\mu$m-sized presolar SiC grains with very high precisions ($<$10\% errors; \citealt{savina,stephan}). The majority ($\gtrsim$ 90\%) of presolar SiC grains identified in extraterrestrial materials, including mainstream, Y, and Z grains (AGB grains hereafter), came from ancient C-rich AGB stars that evolved prior to the Solar System formation \citep{zinner}. Compared to mainstream grains, Y and Z grains are much rarer (~5\% of all SiC each) and their abundances increase with decreasing grain size \citep{zinner,hoppe10,xu15}. Based on their Si and Ti isotopic compositions, Y and Z grains were previously inferred to have originated from low-mass AGB stars of $0.5 \, Z_\odot$ and $0.3 \, Z_\odot$ , respectively \citep{zi07}.
%Y and Z grains, much lower in abundance ($\sim$1-3\% each depending on grain size) compared to mainstream grains ($\sim$90\%), were previously inferred to have originated from low-mass AGB stars of $0.5 \, Z_\odot$ and $0.3 \, Z_\odot$, respectively, based on their Si and Ti isotopic compositions \citep{zi07}. 
However, \citet{liu19} showed that these low-metallicity models adopted by \cite{zi07} cannot consistently explain the Mo isotopic compositions of Y and Z grains and that, more likely, Y and Z grains came from low-mass AGB stars of $\gtrsim 0.7 \, Z_\odot$. A small fraction of presolar SiC, including X and possibly AB grains, also came from ancient Type II supernovae \citep{zinner,liu17_2,hoppe19}. These stellar dust grains were incorporated into small bodies that formed shortly after the Sun's birth. Billions of years later, primitive extraterrestrial materials, fragments of undifferentiated small bodies (e.g. asteroids and, to a lesser extent, comets), fell and delivered their initially incorporated ancient stellar dust on Earth. 
The linkage between these AGB grains and C-rich AGB stars is particularly supported by the \textit{s}-process
isotopic signatures preserved in the grains, which have provided a number of stringent constraints on the nucleosynthesis occurring in AGB interiors (e.g., \citealt{luga03,liu14b,liu15,liu19}). \\
\indent Unfortunately, the distribution (i.e. masses and metallicities) of the parent stars of AGB grains  remains unknown due to the lack of direct observations and thus needs to be inferred based on Galactic evolution models. \citet{gail09} presented a Galactic Chemical Evolution (GCE) model with dust yields computed with synthetic AGB models, based on which they concluded that the majority of presolar SiC grains originated from AGB stars with masses 1.5$\le$ M/M$_\odot \le$4.0 and roughly solar metallicity. More recent studies focusing on the isotopic compositions of AGB grains suggested that AGB grains predominantly came from low-mass stars ($\lesssim$ 3 M$_\odot$) with close-to-solar metallicities (e.g., \citealt{liu18,liu19}) or slightly super-solar metallicity \citep{lewis13}. Finally, \citet{luga18,luga2020} proposed AGB stars with larger masses (M/M$_\odot\sim 3.5-4$) and higher metallicities (Z/Z$_\odot\sim 1.5-2$) as the parent stars of large ($> 1 \mu$m in diameter) mainstream grains. \\
\indent Here we aim at 
%better 
constraining the mass and metallicity distribution of the grains’ parent stars, by coupling chemo-dynamical, stellar and dust growth codes. This will help to discriminate among the proposed \textit{s}-process models currently available in the literature.
\section{Milky Way chemo-dynamical model}\label{secgce}

To quantify the number of AGB stars close to the Sun's birth place at the epoch of its formation, we use the chemo-dynamical model by \citealt{mi13} (hereafter MCM13). This model was created by combining a high-resolution simulation in the cosmological context with a detailed Milky Way chemical evolution model.  The simulation used here was part of a suite of numerical experiments presented by \citet{martig12}, in which the authors studied the evolution of 33 simulated galaxies from redshift z = 5 to z = 0.
To avoid problems with sub-grid physics in fully cosmological chemo-dynamical models (i.e. with on the fly star-formation and chemical enrichment), a detailed semi-analytical chemical evolution model for the Milky Way 
was coupled with the simulation a posteriori, assigning elemental abundances to stars according to their ages, birth radii, and the weighted star-formation history. The unconstrained nature of the simulation allows stars to migrate radially in a fully self-consistent matter. Given the impossibility to follow single stars during their migration, they are grouped in "Star Particles" (hereafter SPs), each with a mass of $7.5\times 10^4$ M$_\odot$. The corresponding hybrid chemo-dynamical model succeeded in reproducing the chemo-kinematic relations in the Milky Way (see \citealt{mi16} and references therein).\\
\indent We expect that stars attaining their AGB phase in the Solar vicinity at the epoch of Sun formation (4.57 Gyr ago) produced the vast majority of presolar SiC grains. 
It is possible that some dust that was produced in AGB stars evolved far from the Sun's birth place, arrived in the Solar vicinity with different delays (on average less than 300 Myr prior to the Sun's birth; \citealt{heck2020}). However, the number of such migrated grains should be very small, due to the large dilution caused by the isotropic emission at the source (matter ejected by low-mass stars has no preferential direction) and to the (probable) destruction caused by the most energetic radiation permeating the ISM. \\
\indent The almost flat age-metallicity relation (AMR) in the Solar neighborhood for the last 5 Gyr (dashed grey curve in Fig. \ref{fig:gce}) seems to suggest that the metallicity distribution of the Galactic disk barely evolved over time\footnote{On the x-axis, time t=0 yr refers to the current date.}.
However, the temporal evolution of the  metallicities of the local stars is also driven by dynamical processes such as stellar migration, which modifies the distribution of stars in the Solar vicinity over time. In fact, the local AMR is largely controlled by long-lived stars, a large fraction of which migrate out from the inner disk, where the metallicity for a given age is higher. This effect is illustrated in Fig. \ref{fig:gce}: the apparent flatness of the observed AMR in the Solar vicinity (HARPS-GTO data) can be explained by the MCM13 model as the result of radial migration and the superposition of stars born at different Galactic radii. As indicated by the color-coded curves in Fig. \ref{fig:gce}, stars born in narrow birth-radius bins have well-defined AMRs. However, when put together, the total sample appears much flatter. This is a case of the statistical phenomenon known as Simpson’s paradox (see \citealt{mi19}).
\begin{figure}[tbp!]
\centering
\includegraphics[width=\columnwidth]{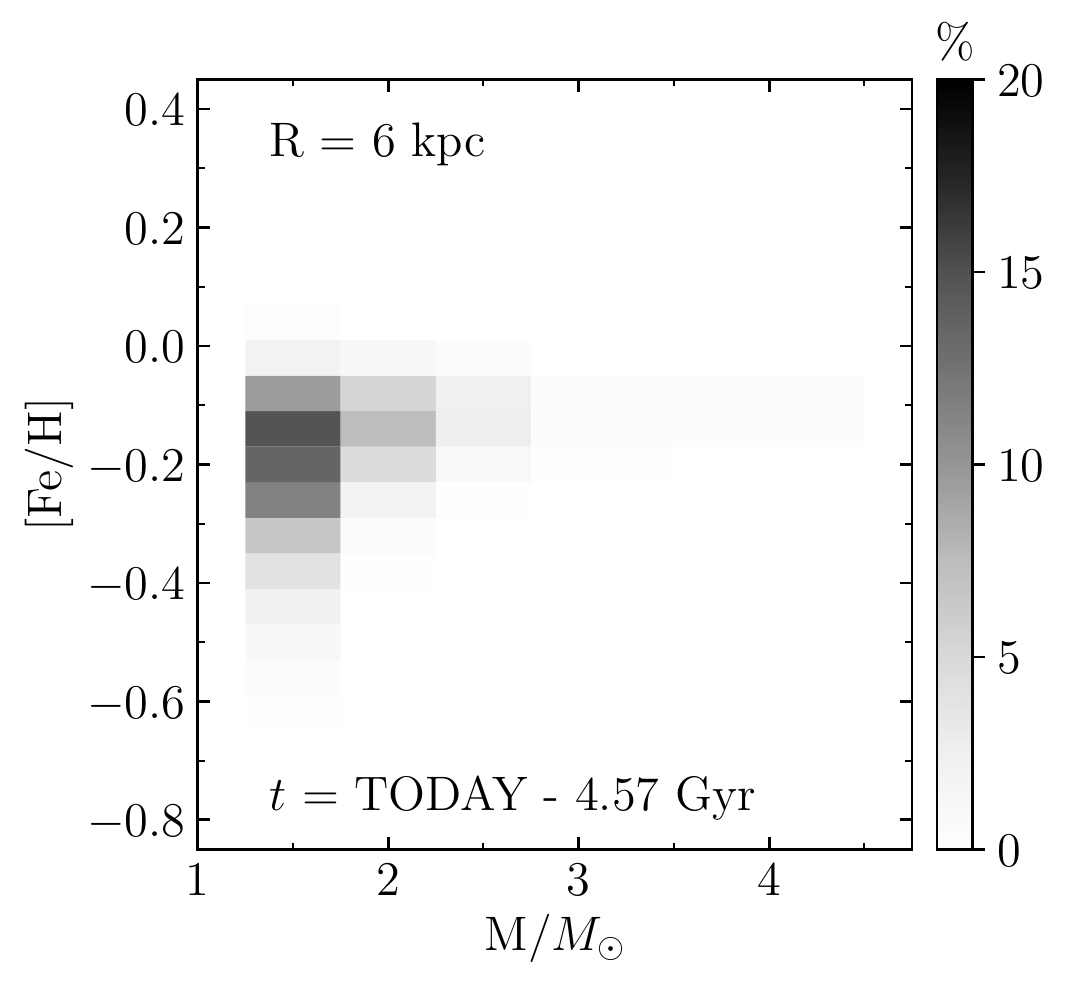}
\caption{Contour plot of the stellar mass-metallicity distribution in the Solar vicinity at the epoch of Solar System formation (MCM13 model output).}
\label{fig:res1}
\end{figure} 
\\
\begin{figure*}[ht!]
\centering
\includegraphics[width=\textwidth]{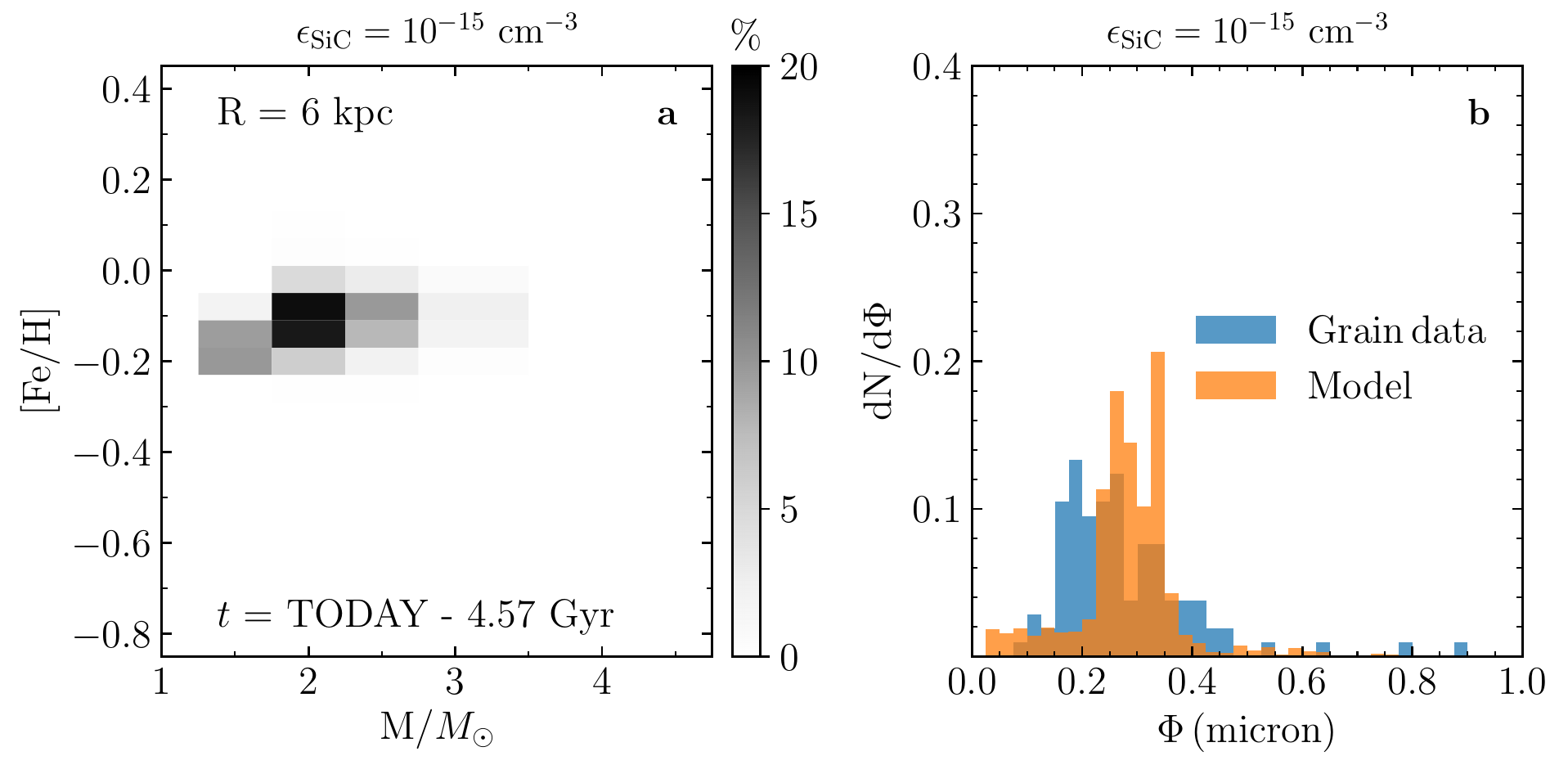}
\caption{Panel (a): same as Fig. \ref{fig:res1}, but with percentages weighted by the SiC grain yields of the corresponding AGB models (initial seed dust number $\epsilon_{\rm SiC}=10^{-15}$ cm$^{-3}$). Panel (b): corresponding  distribution of SiC production versus grain size (in diameter), compared to SiC grains identified in situ in primitive meteorites  (see text for details).}
\label{fig:res_1m15}
\end{figure*} 
\indent The mass-metallicity probability distribution of AGB stars present in the Solar neighborhood at the epoch of its formation (4.57 Gyr ago) obtained with the MCM13 model is shown in Fig. \ref{fig:res1}. 
We assumed that the Sun formed at an inner galactocentric radius (6 Kpc) and subsequently migrated to its current radius of 8 kpc (\citealt{wie96}; \citealt{clayton97}; MCM13). The grid resolution is determined by the resolutions of the masses and metallicities of the available AGB stellar models contained in the FRUITY database\footnote{http://fruity.oa-abruzzo.inaf.it/} \citep{cris09,cris11,pier13,cris15}. In the framework of our AGB models, we did not consider stars with masses below 1.3 M$_\odot$ since they do not experience enough dredge up to become C-rich (except for very low metallicities, which are irrelevant for this study). The MCM13 model was used to directly output the metallicity distribution of considered SPs, in the Solar vicinity at 4.57 Gyr ago, as shown in Fig. \ref{fig:res1}. \\
\indent The determination of the corresponding mass distribution was more complex. In order to obtain the mass distibution, we assume each SP consisting of a simple stellar population. The distribution of SPs in the solar vicinity at the epoch of the Sun formation is already given by the MCM13 model, but we also need to know the number of objects on the AGB phase within each of the SPs, which is intrinsically connected to the SP age. In fact, being a simple stellar population, the mass evolving on the AGB is univocally determined for each SP.
For example, only SPs older than 10 Gyr can host AGB stars with initial masses below 1 M$_\odot$. In order to extract this quantity, we grouped SPs in temporal bins, which are defined based on the timescale needed for a star to reach the AGB (depending on the initial stellar mass). For instance, a 1.5 M$_\odot$ star needs approximately 3 Gyr to attain the AGB phase, while a 6.0 M$_\odot$ only needs about 70 Myr. Given that we know the distribution of SPs in each temporal bin, in order to extract the number of AGB stars from each SP, we applied a Salpeter initial mass function (IMF; \citealt{salpeter}).\\
%Our goal is to determine the fraction of {\bf star} particles hosting stars in their AGB phase for a given mass interval. As a matter of fact, the initial mass sets an equivalent timescale needed for the star to attain its AGB phase. For example, only stellar particles older than 10 Gyr can host AGB stars with initial masses below 1 M$_\odot$.  
%In detail, we calculate the number of stellar particles born in the corresponding time interval for each selected mass range, based on which we derived the number of stellar particles hosting stars in that mass range at 4.57 Gyr ago. 
%Then, we apply  a Salpeter initial mass function (IMF; \citealt{salpeter}) to determine the number of AGB stars in these stellar particles, each of which consists of a simple stellar population.\\
\indent The predicted distribution in Fig. \ref{fig:res1} is largely weighted toward 1.5 M$_\odot$ mass at lower-than-solar metallicities (Z $\simeq 0.7$ Z$_\odot$)\footnote{In FRUITY models, Z$_\odot=1.4\times 10^{-2}$.}. 
This is a direct consequence of the IMF peaking at extremely low masses. Finally, to investigate their contributions to the presolar SiC inventory in the Solar System, these AGB stars in Fig. \ref{fig:res1} have to be further weighted by their SiC dust production.

\section{Dust growth model}
\label{secsic}

In order to determine SiC yields in AGB stars, we adopted a dust growth code, coupled with wind dynamics, initially based on \citet{fg06}, but with some important modifications in the input physics \citep{nanni13,nanni16,nanni19}. For each mass-metallicity combination identified by the MCM13 model, we calculated the corresponding dust yield. The reference stellar evolutionary models are taken from the FRUITY database. For each model we extract physical inputs along the AGB track (i.e. luminosity, current stellar mass, temperature, mass-loss rate and surface chemical abundances). As output, we obtain the chemical composition of the produced dust, the relative yields of different dust species and the expansion velocity of the external layers of the circumstellar envelope.\\
\indent In this work we introduced a time-averaged shock density profile in the inner part of the envelope before the onset of the dust-driven wind, starting from the prescriptions by \citet{che92}. Such a modification is necessary for reliably modelling the condensation of SiC. Indeed, SiC is condensed at higher temperatures with respect to carbon (amorphous) dust. The formation of such a dust species, under favorable conditions, accelerates the outflow via dust-driven wind. In particular, the original stationary wind profile underestimates the density at the formation radius of SiC, making it difficult to produce large SiC grains (highlighted by laboratory measurements; see Next Section). \\
%In order to determine SiC yields in AGB stars, we adopted the dust growth code by \citet{nanni13,nanni14}. We improved the stationary wind model by including a shock time-averaged density profile in the dust formation zone and an updated temperature profile in the innermost zone. 
\indent Operatively, in order to describe the circumstellar envelope, we ideally divide it into three spatially separated regions: a) from the photosphere to the condensation radius, $R_{\rm cond}$, where the first dust species form; b) from $R_{\rm cond}$ to the radius at which the outflow starts to accelerate through dust-driven wind, $R_{\rm acc}$; c) from $R_{\rm acc}$ outwards.
For the temperature profile from $R_{\rm cond}$ outwards, we adopt the Lucy approximation \citep{lucy76} as already used in several other works \citep[e.g.][]{fg06,ventura12,nanni13}.
For describing the inner temperature profile from the photosphere to $R_{\rm cond}$, we adopt the following temperature profile:
    \begin{equation}\label{Eq:Tinn}
        T_{\rm inner}(r)=T_{\rm eff} \Big(\frac{r}{R_*}\Big)^{-\alpha_{\rm T}},
    \end{equation}
where $T_{\rm eff}$ is the effective temperature, $R_*$ the stellar radius and $\alpha_{\rm T}$ is a parameter determined by linking the inner and outer temperature profiles at $R_{\rm cond}$.\\
\indent For describing the density profile in the region with dust-driven wind in region c), the stationary wind equation is adopted \citep[e.g.][]{fg06,ventura12,nanni13}.
For describing the density profile in the regions a) and b), we adopt the time-averaged approximation for the shock extended zone presented in \citet{che92}:
\begin{equation}\label{Eq:rho_inn}
        \rho(r)=\rho_0\times \exp{\int_{R*}^{r} -\frac{(1-\gamma_{\rm shock}^2)}{H_0(r^{\prime})}dr^{\prime}}, 
\end{equation}
where R$_*$ is the photospheric radius, $\rho_0$ is the value of the density at the photosphere, $\gamma_{\rm shock}$ is the shock strength and is evaluated by linking the inner density profile with the density value obtained at $R_{\rm acc}$ from the stationary wind profile. The quantity $H_0(r)$ is given by:
\begin{equation}\label{Eq:H0}
    H_0(r)=\frac{kT(r)r^2}{\mu m_{\rm H} G M_*},
\end{equation}
where $k$ is the Boltzmann constant, $T(r)$ the temperature profile, $\mu$ is the mean molecular weight, $m_{\rm H}$ the mass of hydrogen atom, $G$ the gravity constant, and $M_*$ the stellar mass.
The density profile is numerically evaluated by performing the integral of $H_0(r)$ in Eq. \ref{Eq:rho_inn}. We note that we assumed a different temperature profile with respect to \citet{che92}, who choose a fixed value for the parameter $\alpha$=0.6   (assumed representative for describing the temperature profile of the circumstellar envelope of the carbon star IRC+10216). Since in our work we need to simulate a variety of carbon stars evolving through the AGB phase, the adoption of the same kind of temperature profile is not straightforward, because we would need to predict
how $\alpha$ changes as a function of the stellar parameters. Therefore, we adopted the temperature
profile by \citet{lucy76} from $R_{\rm cond}$ outwards, which is consistently computed taking into account the amount of dust formed in the circumstellar envelope as a function of the stellar parameters. Then, in
order to compute the integral of $H_0(r)$ in Eq. \ref{Eq:rho_inn}, we adopt the temperature profile from the photosphere to $R_{\rm cond}$ given by Eq. \ref{Eq:Tinn}, consistently matched with the outer temperature profile.
\\
%A brief outline of this scheme is presented in Appendix~\ref{appen}.
%The most important dust species in C-rich circumstellar envelopes is amorphous carbon. 
%The calibration of the free parameters involved in its growth process and optical properties has been presented in \citet{nanni16} (see also \citealt{nanni19}): we report a short summary in Appendix~\ref{appen2}. Hereafter, we focus on the growth process of SiC dust.
\indent 
%In addition to O-rich dust (olivine, pyroxene, corundum, quartz, periclase, metallic iron),
We model the grain growth process on starting seed nuclei of two types of carbonaceous grains: amorphous carbon and SiC. For amorphous carbon, the seed particle abundance $\epsilon_{\rm C}$ - i.e. the abundance of particles, with respect to hydrogen atoms, on which other molecules accrete - was selected in order to yield typical dust grains of size $\lessapprox 0.04$ $\mu$m. The value of $\epsilon_{\rm C}$ is proportional to the carbon excess: $\epsilon_{\rm C}=\epsilon_{\rm C,0}\times (C-O)$. The value of $\epsilon_{\rm C,0}$ together with optical constants from \citet{hanner88} have been selected to reproduce both the photometry in the near and mid-infrared bands \citep{nanni16} and Gaia Data Release 2 \citep{nanni19}.
The condensation of amorphous carbon is assumed to occur by a successive addition of C$_2$H$_2$ molecules to the seed particle below a temperature of 1100 K \citep{ff89,
che92}. The probability for molecules (in our case C$_2$H$_2$) to stick on the grain surface is known as ``sticking coefficient''. Those coefficients cannot be derived from theoretical calculations and are poorly constrained by laboratory experiments. We initially adopt a sticking coefficient equal to unity for amorphous carbon (as well as for SiC, see below). However, with such a choice we were not able to reproduce the observed expansion velocity vs mass-loss rates observed for Galactic carbon-stars. As already highlighted, carbon dust is responsible for the outflow acceleration: a sticking coefficient equal to unity would lead to an outflow expansion velocity larger than the observed ones. On the other hand, a sticking coefficient of 0.2 reduces the expansion velocity by lowering the amount of amorphous carbon and, coupled to an initial wind speed of 4 km s$^{-1}$, produces final velocity profiles that are consistent to observations. \\
%The observed wind speed as a function of the mass-loss rate is compared with our model predictions. In order to reproduce such an observable, we selected an initial wind speed of 4 km s$^{-1}$ and a , equal to 0.2. 
\indent SiC formation is assumed to proceed through the addition of Si atoms and C$_2$H$_2$ molecules on the seed. We maintain the original value of a sticking coefficient for SiC equal to unity. A variation in the sticking coefficient for SiC does not affect the outflow expansion velocity, since the formation of SiC is not responsible for accelerating the outflow. We did not find valid reasons to modify the SiC sticking coefficient, especially considering the lack of observational constraints. The condensation of SiC is prevented at high temperatures by chemisputtering, i.e. the destructive bombardment of a solid by molecular hydrogen (H$_2$). Below a certain temperature, which is model-dependent, destruction becomes inefficient and SiC grains can thus grow unhindered. We adopted the optical constants from \citet{pitman} for $\beta$-SiC, since structural analyses of presolar SiC grains revealed that they consist dominantly of $\beta$-SiC, also known as cubic SiC (3C-SiC; \citealt{daulton,liu171}). For the seed particle abundance, we choose $\epsilon_{\rm SiC}=10^{-15}$. In order to evaluate the sensitivity of our results to this free parameter, we also tested significantly larger ($\epsilon_{\rm SiC}=10^{-13}$) and lower ($\epsilon_{\rm SiC}=10^{-18}$) values (see next Section). We expect some degeneracy between the results obtained with a large sticking coefficient and a low seed particle abundance and the results obtained with a low sticking coefficient and a large seed particle abundances. For this reason we varied one parameter only ($\epsilon_{\rm SiC}$), which, in any case, produces the largest differences as far as the total predicted mass of SiC dust is concerned.\\

\begin{figure*}[ht!]
\centering
\includegraphics[width=\textwidth]{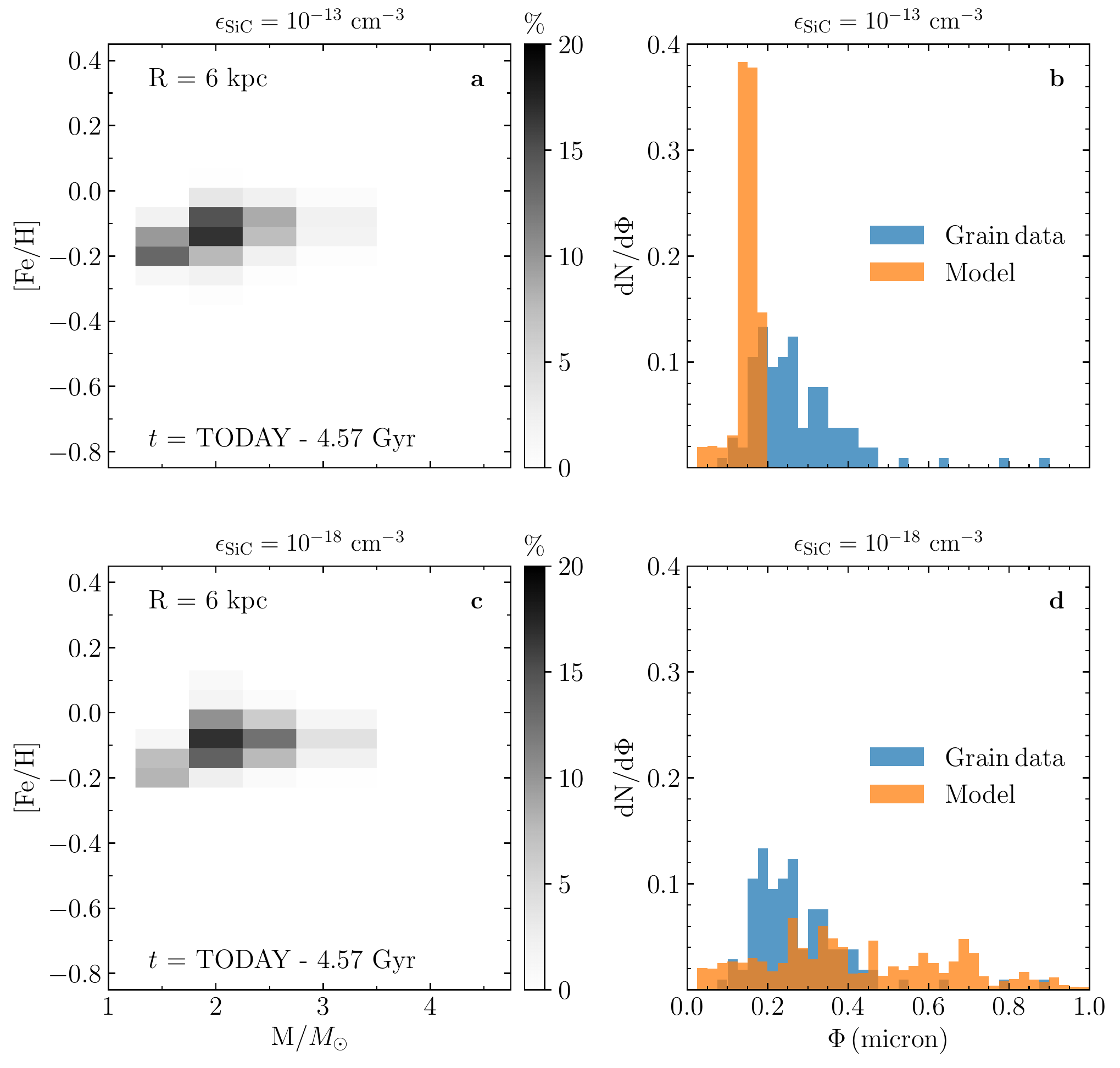}
\caption{Same as Fig. \ref{fig:res_1m15}, but assuming $\epsilon_{\rm SiC}=10^{-13}$ cm$^{-3}$ (panels a and b) and $\epsilon_{\rm SiC}=10^{-18}$ cm$^{-3}$ (panels c and d) See text for details. }
\label{fig:res_1m13_1m18}
\end{figure*} 

\section{Results and Discussions}

Once the SiC dust yields were calculated based on FRUITY models, these values were used to assign weights to the percentages obtained with the chemo-dynamical MCM13 model (Fig. \ref{fig:res1}). The weighted SiC production distribution %($\epsilon_{\rm SiC}=10^{-15}$ cm$^{-3}$) 
is shown in Fig. \ref{fig:res_1m15}a, which reveals a number of interesting features as summarized below:
\begin{enumerate}
\item{Since this study focuses on SiC dust, only C-rich models were considered in our simulations. This led to a reduced number of considered mass and metallicity combinations. As a result, the 1.5 M$_\odot$ stars at large metallicities and, in general, stars more massive than 3 M$_\odot$ are thus excluded in our simulations, because they are predicted to be oxygen-rich by FRUITY stellar models. Although this prediction depends strictly on the Third Dredge Up (TDU) efficiency of our stellar models, the number of these stellar objects is negligible in the Solar neighborhood at 4.57 Gyr ago according to the MCM13 modelling result. In other words, our results shown in Fig. \ref{fig:res_1m15} are barely affected by these stellar objects even if they could become carbon-rich.} 
\item{Fig. \ref{fig:res_1m15}a illustrates that $\sim$ 2 M$_\odot$ AGB stars are the dominant contributors to AGB grains in the Solar System, resulting from the fact that they were numerous at the epoch of its formation and are predicted to dredge up more material with respect to lower-mass stars (see, e.g., \citealt{cris11}).}
\item{Fig. \ref{fig:res_1m15}a reveals a general shift to larger metallicities with respect to Fig. \ref{fig:res1}. This can be easily understood in terms of the SiC formation process, which is governed by carbon and silicon abundances. While the former depends on the TDU efficiency, the latter is directly proportional to the metallicity of the model. Thus, the higher the metallicity, the higher the expected SiC yield. It is noteworthy that an extra-production of $^{12}$C, which occurs at lower metallicities, does not directly translate to a higher production of SiC, because the carbon excess tends to form extra amorphous carbon (see \citealt{nanni13} and references therein).}
\end{enumerate}
By extracting presolar SiC grains from Murchison meteorite via acid dissolution, \citet{amari94} studied the grain size distribution and found that the grain size distribution has a maximum at 0.4 $\mu$m, and follows a power law with an exponent between -4 and -5 for sizes between 0.7 and 3.2 $\mu$m. This estimate, however, is likely skewed toward larger sizes, since the extraction of 
smaller grains is expected to be less efficient. In addition to the acid-extraction method, presolar SiC grains can also be directly identified in situ in primitive meteorites by isotopic mapping. Such samples should better represent the whole population of AGB grains incorporated during the Solar System formation. We calculated the size distribution of AGB grains identified in situ in the most primitive meteorites \citep{floss,ngu,nit18,hae18} to eliminate the effect of parent-body processing on varying the grain size \citep{davi14}. The result (blue histogram) is compared to our model predictions (orange histogram),  in Fig. \ref{fig:res_1m15}b. The model result overlaps with the grain data with a small shift to larger grains (by $\sim$100 nm).\\
\indent In order to strengthen our results, we tested  larger and lower initial seed particle abundances ($\epsilon_{\rm SiC}=10^{-13}$ and $10^{-18}$ cm$^{-3}$, in comparison to $\epsilon_{\rm SiC}=10^{-15}$ cm$^{-3}$ in Fig. \ref{fig:res_1m13_1m18}) in our calculations. The results are shown in Fig. \ref{fig:res_1m13_1m18}b and \ref{fig:res_1m13_1m18}d. The $\epsilon_{\rm SiC}=10^{-13}$ cm$^{-3}$ case matches the small-size tail of the grain distribution.
%, even if the latter should be considered as an upper limit. In fact, the technique used to identify SiC grains in situ has a spatial resolution of 100-150 nm, such that a number of smaller grains were probably missed during the search \citep{hoppe17}. 
%
% \begin{figure*}[ht!]
% \centering
% \includegraphics[width=\textwidth]{n_H_1e-18.pdf}
% %\gridline{\fig{INI_6Kpc_NORM_n_H_1e-18g.pdf}{0.54\textwidth}{(a)}
% %          \fig{average_1m18.pdf}{0.46\textwidth}{(b)}}
% \caption{Same as Fig. \ref{fig:res_1m15}, but assuming $\epsilon_{\rm SiC}=10^{-18}$ cm$^{-3}$ (see text for details). }
% \label{fig:res_1m18}
% \end{figure*} 
%
On the other hand, the $\epsilon_{\rm SiC}=10^{-18}$ cm$^{-3}$ case predicts the formation of large grains, a consistent fraction of which ranges from 0.5 to $\sim$1 $\mu$m in size (thus larger than grain data). Thus, $\epsilon_{\rm SiC}=10^{-18}$ cm$^{-3}$ can be considered as a lower limit for the seed particle abundance. The corresponding mass-metallicity distributions of the AGB stars are reported in Fig. \ref{fig:res_1m13_1m18}a and Fig. \ref{fig:res_1m13_1m18}c. Based on Fig. \ref{fig:res_1m15} and \ref{fig:res_1m13_1m18}, we conclude that AGB stars with M $\sim$ 2 M$_\odot$ and Z $\sim$ Z$_\odot$ are the dominant contributors to AGB grains identified in primitive extraterrestrial materials. The predicted distributions of AGB stars in Figures \ref{fig:res_1m15}a, and \ref{fig:res_1m13_1m18}a and \ref{fig:res_1m13_1m18}c also provide a natural explanation to the abundances and isotopic compositions of Y and Z grains. Y and Z grains were constrained to have come from low-mass AGB stars of $\gtrsim 0.7 \, Z_\odot$ based on comparison of their Mo isotopic compositions with FRUITY model predictions \citep{liu19}. Both the lower limit of their parent star metallicities and their low abundances are consistent with the decreasing probabilities of low-mass AGB stars toward lower metallicities ([Fe/H]=-0.2, corresponding to 0.63 $Z_\odot$, represents a sort of lower limit of our metallicity distribution).
Note that the metallicity distribution only varies slightly from $\epsilon_{\rm SiC}=10^{-13}$ cm$^{-3}$ to $\epsilon_{\rm SiC}=10^{-18}$ cm$^{-3}$, thus demonstrating that our conclusions are barely affected by the choice of $\epsilon_{\rm SiC}$ (which is by far the largest uncertainty source of the SiC dust growth process). As a matter of fact, we demonstrated that the distribution of parent AGB stars of presolar grain is dominantly shaped by the result of the MCM13 model.\\
\indent Besides the differences in the physical recipes of the dust growth models (see the discussion in Section \ref{secsic}), our conclusions are similar to the results by \citet{gail09}. Nonetheless, some differences have to be highlighted. In particular, in the study of \citet{gail09}, parent AGB star distribution is characterized by lower metallicities and larger masses. First, let us note that a standard GCE model (as the one adopted by \citealt{gail09}) cannot obtain super-solar metallicities at the moment of the formation of the solar system. This is an intrinsic property of one-dimension GCE models, in which the mean metallicity is expected to grow with time. In contrast, with our chemo-dynamical model we found that metal-rich stars ($Z>Z_\odot$) may migrate from the inner galaxy to the galacto-centric position where the Sun formed. This automatically reduces the relative contribution from low metallicity AGBs. Moreover, due to the long migration timescales of these stars, they also affect the mass distribution of SiC parent stars. As a matter of fact, low mass (1.5-2.0 M$_\odot$) metal-rich stars are favored, because they evolve on long evolutionary timescales (and thus they have time to migrate). Our mass distribution is weighted toward lower initial stellar masses also because of a different TDU efficiency. In their synthetic approach, \citet{gail09} adopted the TDU efficiencies of AGB models by \citet{kakka02}. Those models are characterized by a larger TDU efficiency in the 3-4 M$_\odot$ mass range with respect to our models (for a comparison see \citealt{cris11}), resulting in the larger weight of 3-4 M$_\odot$ in the final SiC parent AGB star distribution with respect to our results. \\
\indent Recently, \citet{luga2020} (see also \citealt{luga18}) proposed that presolar SiC grains originated from more massive (M $\sim$ 4 M$_\odot$) and more metal-rich ($Z \sim  2 \times Z_\odot$) AGB stars. %However,
It is noteworthy that the grains discussed in these papers are unusually large ($>1 $ $\mu$m) and correspond to %only 
a very minor fraction of presolar SiC, in contrast to the much smaller grains discussed here, which are the dominant population of AGB grains. Thus, the discrepancy could be resolved, if the grain size is metallicity-dependent, i.e., decreasing grain size with decreasing metallicity. Note that this hypothesized size-metallicity correlation seems to be supported by the size-dependent Sr and Ba isotopic anomalies observed in bulk meteoritic SiC acid residues (\citealt{ott,podo}). In this scenario, the small population of large AGB grains is well explained by the rarity of super-solar metallicity AGB stars present in the Solar vicinity at 4.57 Gyr ago, as predicted by the MCM13 model. In comparison, $\sim$2 M$_\odot$, $\sim$Z$_\odot$ AGB stars were much more common, resulting in the dominant population of small grains, as shown in Fig. \ref{fig:res_1m15} and \ref{fig:res_1m13_1m18}. 
%It is noteworthy that the proposal of 4 M$_\odot$, 2 Z$_\odot$ AGB stars as the parent stars of AGB grains \citep{luga18,luga2020} was derived mainly based on grain-model comparison for heavy-element isotopic compositions. 
%In contrast to the result of \citet{luga18,luga2020}, 
Interestingly, \citet{vesc20} recently showed that the same set of measured AGB grain data can be well explained by magnetic FRUITY model predictions for 2 M$_\odot$ AGB stars with close-to-solar metallicities, thus supporting our result.\\
%, thus questioning the robustness of the Lugaro et al. proposal. \\
\indent Finally, we would also like to clarify the explanation to the recent proposal by \citet{ek2020} that the Solar System {\it s}-process component was produced with a higher efficiency than the {\it s}-process component recorded in bulk AGB grains. First of all, we note that this proposal in fact is a natural consequence of GCE. During GCE, gas, the dominant form of ISM material from which the Solar System formed, preserved all its memory of {\it s}-process production over the previous $\sim$6.5 Gyr, while the incorporated AGB grains mostly recorded the {\it s}-production from the last $\sim$300 Myr prior to the Solar System formation \citep{heck2020}. Since the efficiency of the {\it s}-process increases with decreasing metallicity and in turn decreases over time due to the AMR, the Solar System thus naturally contains a more efficient {\it s}-process component than its incorporated AGB grains. 
However, this natural difference between the Solar System and its incorporated AGB grains may not be the explanation to the small Pd isotopic anomalies with respect to Mo and Ru observed by \citet{ek2020}. This is because the Solar System abundances for the three elements received only $\sim$35\% to $\sim$50\% contributions from the {\it s}-process \citep{pra2020}, the missing part coming from the {\it r}-process. The latter may have enhanced their Solar System abundances with different efficiencies, as it is expected in astrophysical {\it r}-process sites with varying electron abundances (e.g., production of higher-than-solar Mo/Pd and Ru/Pd by the {\it r}-process with $Y_e$ of 0.25-0.30; see, e.g., \citealt{lippuner}). 
Hence, %in contrast to the rationale given by \citet{luga2020}, 
the smaller-than-expected Pd isotopic anomalies observed by \citet{ek2020} does not necessarily suggest that AGB grains dominantly came from super-metallicity AGB stars and is irrelevant when discussing the absolute metallicities of the parent stars of AGB grains.
\section{Conclusions and future plans}\label{conclu}
We determined the mass and metallicity distribution of presolar SiC parent stars, by integrating a numerical Milky Way chemo-dynamical model with dust yields from FRUITY models. Our models predict that the SiC production at the epoch of the Solar System formation is dominated by contributions from AGB stars with M $\sim$ 2 M$_\odot$ and Z $\sim$ Z$_\odot$, which are thus likely the parent stars of most presolar SiC grains identified in extraterrestrial materials. \\
\indent In the future, we plan to investigate the proposed metallicity-size correlation (based on size-dependent isotopic anomalies observed in bulk SiC acid residues) by implementing our treatment of magnetic buoyancy \citep{vesc20} in all FRUITY stellar models. As shown by \citet{vesc20}, such a treatment is needed for FRUITY models to reproduce the {\it s}-process isotopic signatures of AGB grains. By adopting magnetic FRUITY stellar models in the framework of the analysis presented in this paper, we will be able to provide further insights into the suggested metallicity-size correlation.\\ \indent Finally, it is noteworthy that there still exist a number of unknowns in  AGB modelling, such as the efficiency of TDUs, the mass-loss law, and the origin of mixing processes at work in stellar interiors. Dust nucleation (theory), describing the transition from gas phase molecules to dust grains via molecular clusters, appears even more uncertain \citep{david17}. Unlike their bulk analogue ($\beta$-SiC), the smallest SiC clusters are characterized by atomic segregation of silicon and carbon atoms. It is still a matter of debate, at which cluster size SiC favours structures with alternating Si-C atomic ordering since the SiC cluster energies strongly depend on the employed level of theory (density functional).
%Dust nucleation (theory), directly correlated to molecular clusters, appears even more uncertain regarding the transition from gas phase species to dust grains \citep{david17}. 
For this reason, further theoretical investigation is required, as well as a corresponding thorough experimental verification.
\begin{acknowledgements}
We thank the anonymous referee for a quick, detailed and helpful report, which largely improved the quality of our paper. A.N. acknowledges the support of the Centre National d’Etudes Spatiale (CNES) through a post-doctoral fellowship. G.C. acknowledges financial support from the EU COST Action CA16117 (ChETEC). N.L. acknowledges financial support from NASA (80NSSC20K0387 to N.L.). D.G. acknowledges support from the ERC consolidator grant 646758 AEROSOL.
\end{acknowledgements}
\bibliographystyle{aa} 

\end{document}